\documentclass[letter, traditabstract]{aa} 
\usepackage{graphicx}
\usepackage{txfonts}
\newcommand{\OI}{O\,{\sc i}}

\newcommand{\HII}{H\,{\sc ii}}

\begin{document} 
   \title{Spatially resolved images of reactive ions in the Orion Bar\thanks{This paper makes use of the following ALMA data: ADS/JAO.ALMA\#2012.1.00352.S.  ALMA is a partnership of ESO (representing its member states), NSF (USA), and NINS (Japan), together with NRC (Canada), and NSC and ASIAA (Taiwan), in cooperation with the Republic of Chile. The Joint ALMA Observatory is operated by ESO, AUI/NRAO, and NAOJ.}$^,$\thanks{Includes IRAM\,30\,m telescope
    observations. IRAM is supported by INSU/CNRS (France), MPG (Germany), and IGN (Spain).}}

   \author{Javier R. Goicoechea\inst{1}
          \and
          Sara Cuadrado\inst{1}
          \and
          J\'er\^{o}me Pety \inst{2,3}
          \and
          Emeric Bron \inst{1,3}
          \and
          John H. Black \inst{4}
          \and\\
          Jos\'e Cernicharo \inst{1}
          \and
          Edwige Chapillon \inst{2,5}
          \and
          Asunci\'on Fuente \inst{6}
          \and
          Maryvonne Gerin \inst{3}
          }

\institute{Instituto de Ciencias de Materiales de Madrid (CSIC), 28049, Madrid, Spain. \email{jr.goicoechea@icmm.csic.es}
\and
 Institut de Radioastronomie Millim\'etrique, 38406, 
 Saint Martin d’H$\grave{\rm e}$res, France.
\and
  LERMA, Obs. de Paris, PSL Research University, CNRS, Sorbonne Universite\'es, 
  UPMC Univ. Paris 06, ENS, F-75005, France.     
\and
  Chalmers University of Technology, Onsala Space Observatory, 43992 Onsala, Sweden.
\and
OASU/LAB-UMR5804, CNRS, Universite\'e Bordeaux, 33615 Pessac, France.
\and
Observatorio Astron\'omico Nacional (IGN). Apartado 112, 28803 Alcal\'a de Henares, Spain.
}

\date{Received 1 March 2017 /  Accepted 21 April 2017}

  \abstract{
We report high angular resolution (4.9$''$$\times$3.0$''$) images of  reactive ions SH$^+$, HOC$^+$, and SO$^+$ toward the Orion Bar photodissociation region (PDR). We used ALMA-ACA to map  several rotational lines at 0.8\,mm, 
 complemented with multi-line observations 
obtained with the IRAM\,30\,m telescope.
 The SH$^+$ and HOC$^+$ emission is restricted to a narrow layer of 2$''$- to 10$''$-width \mbox{($\approx$800 to 4000\,AU} depending on the assumed PDR geometry) that follows the vibrationally excited H$_{2}^{*}$ emission. Both ions efficiently
  form very close to the \mbox{H/H$_2$} transition zone,
  at a depth of $A_{\rm V}$$\lesssim$1\,mag into
the neutral cloud, where abundant C$^+$, S$^+$, and H$_{2}^{*}$ coexist. SO$^+$ peaks slightly deeper into the cloud. The observed  ions have low rotational temperatures   
(\mbox{$T_{\rm rot}$$\approx$10-30\,K$\ll$$T_{\rm k}$}) and narrow line-widths  \mbox{($\sim$2-3\,km\,s$^{-1}$)}, a factor of $\simeq$2 narrower that those of the lighter  reactive ion CH$^+$. This is consistent with the higher reactivity and faster radiative pumping rates of  CH$^+$  compared to the heavier ions, which are driven relatively faster toward smaller velocity dispersion by elastic collisions and toward lower $T_{\rm rot}$ by inelastic collisions. We estimate column densities and  average physical conditions  
 from an excitation  model  (\mbox{$n$(H$_2$)$\approx$10$^5$-10$^6$\,cm$^{-3}$}, \mbox{$n(e^-)$$\approx$10\,cm$^{-3}$}, and \mbox{$T_{\rm k}$$\approx$200\,K}). 
Regardless of the excitation details,  SH$^+$ and HOC$^+$  
clearly trace the most exposed layers of the UV-irradiated molecular cloud surface, whereas SO$^+$ arises from slightly more shielded layers.
}

   \maketitle
%

\section{Introduction}

Reactive  ions  are transient species for which the timescale of reactive collisions with H$_2$, H, or $e^-$ (leading to a chemical reaction, and thus molecule destruction) is comparable to, or shorter than, that of inelastic collisions \mbox{\citep[][]{Black_1998,Nagy_2013,Godard_2013}}.
The formation of  reactive ions such as CH$^+$ and SH$^+$ 
 depends on the availability of C$^+$ and S$^+$ (i.e.,~of UV photons
 and thus high ionization fractions $x_e$=$n(e^-)/n_{\rm H}$), and on the presence of excited H$_2$ (either \mbox{UV-pumped} or hot and thermally excited). 
This allows overcoming the high endothermicity (and sometimes energy barrier) of some of the key initiating chemical reactions \citep[e.g.,][]{Gerin_2016}. The reaction \mbox{C$^+$ + H$_2$(v) $\rightarrow$ CH$^+$ + H}, for example, is
endothermic by \mbox{$\Delta E/k\simeq 4300$\,K} if v=0, but exothermic and fast for v$\geq$1 \citep{Hierl_1997,Agundez_2010}. Despite their short lifetimes,
reactive ions can be detected and used to probe energetic processes in irradiated, circumstellar \citep[e.g.,][]{Cernicharo_1997},  interstellar \citep[e.g.,][]{Fuente_2003}, or protostellar \citep[e.g.,][]{Benz_2016} gas.

\begin{figure*}[ht]
\centering
\includegraphics[scale=0.57,angle=0]{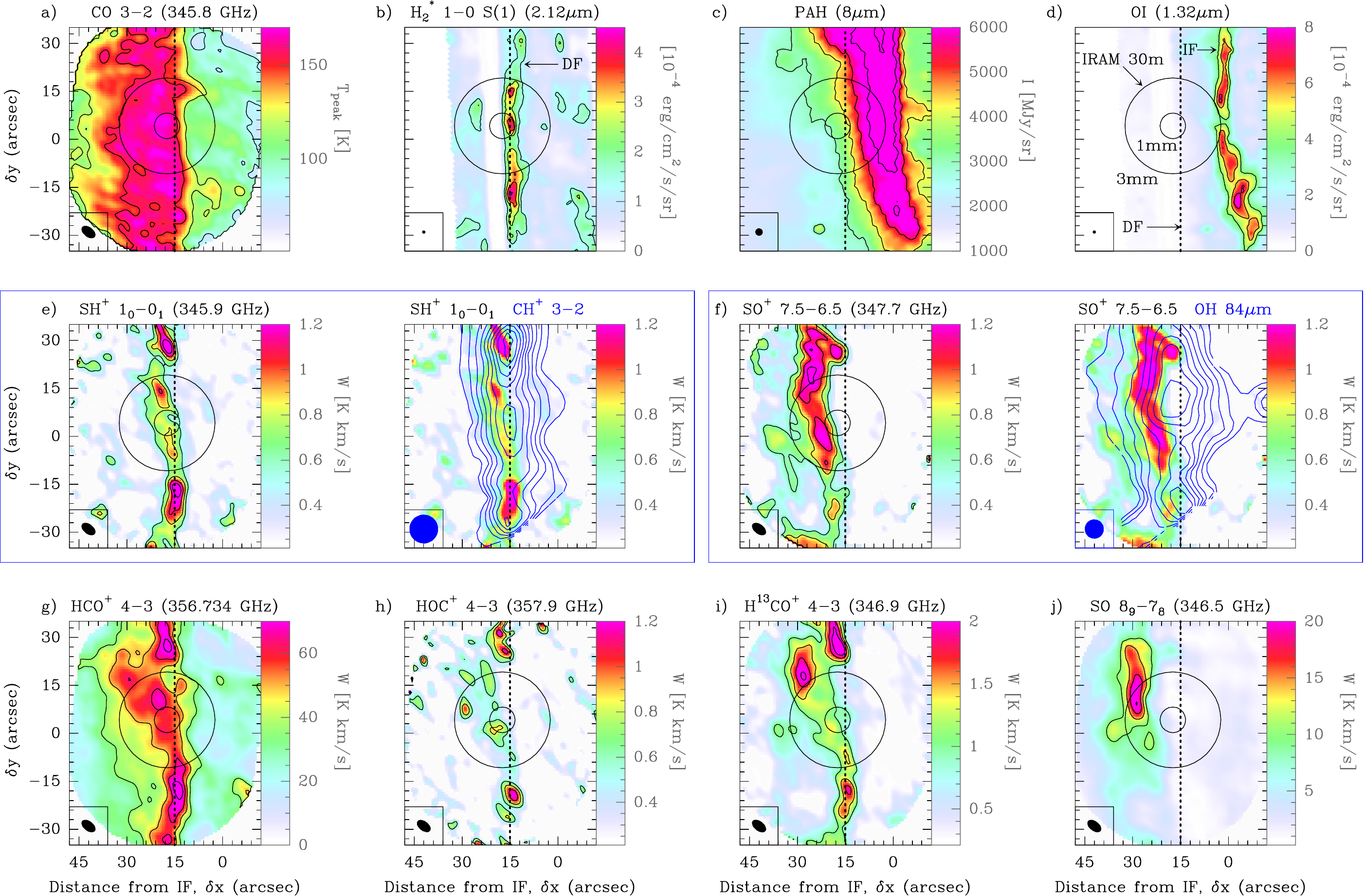}
\caption{ALMA-ACA and complementary images of the Bar.  All
images have been rotated to bring the FUV illuminating direction roughly in the horizontal direction (from the right). 
The circles indicate the position of our IRAM\,30\,m line survey, with beams
of $\approx$8$''$ ($\approx$30$''$) at 1\,mm (3\,mm).
The upper row shows images of b) the H$_{2}$ v=1-0 $S$(1) line
at 2.12~$\mu$m, delineating the DF \citep{Walmsley_2000},  c) the Spitzer\,8\,$\mu$m emission produced mainly by PAHs \citep{Megeath_2012},
and d) the fluorescent \OI~line at 1.32~$\mu$m at the \HII/PDR boundary \citep{Walmsley_2000}. 
The blue contours in e) and f) represent the CH$^+$~119.8\,$\mu$m ($\approx$9$''$ resolution) and OH 84.6\,$\mu$m ($\approx$6$''$) 
lines mapped by \textit{Herschel} \citep{Parikka_2017}.}
\label{fig_figuron}
\end{figure*}

CO$^+$ and HOC$^+$ (the metastable isomer of HCO$^+$) have been
detected in low angular resolution observations of  \mbox{UV-irradiated} clouds near massive stars \citep[][]{Stoerzer_1995,Apponi_1999,Fuente_2003}. They
are predicted to form close to the H/H$_2$ transition zone, the dissociation front (DF), by  high-temperature reactions of C$^+$ with OH and H$_2$O, respectively 
\mbox{\citep[e.g.,][]{Sternberg_1995}}. HOC$^+$~also forms by the
reaction \mbox{CO$^+$ + H$_2$ $\rightarrow$ HOC$^+$ + H}; thus,  CO$^+$ and HOC$^+$ abundances are likely related \citep[][]{Smith_2002}.
In photodissociation regions (PDRs), SO$^+$ is predicted to form primarily 
via the reaction \mbox{S$^+$ + OH $\rightarrow$ SO$^+$ + H} \citep[e.g.,][]{Turner_1996}.
These ion-neutral reactions leading to CO$^+$, HOC$^+$ and SO$^+$ are highly exothermic.

\mbox{\textit{Herschel}}  allowed the 
detection of OH, CH$^+$, and SH$^+$ emission  toward dense PDRs   \citep{Goicoechea_2011,Nagy_2013,Nagy_2017,Parikka_2017}. 
Unfortunately, the limited size of the space telescope did not permit us to resolve
the $\Delta$A$_V$$\lesssim$1\,mag extent of the DF
(a few arcsec for the closest PDRs). Therefore, the
true spatial distribution of the reactive ions emission is mostly unknown.
Unlike CH$^+$,  rotational lines
of SH$^+$ can be observed from the ground \citep{Muller_2014}. 
Here we report the first interferometric images of
SH$^+$, HOC$^+$, and SO$^+$. 

The Orion Bar is a dense PDR \citep[][]{Hollenbach_1999} 
illuminated by a far-UV (FUV) \mbox{(6\,eV$<E<$13.6\,eV)} field  of a few 10$^4$ times
the mean interstellar radiation field. Because of its proximity 
\citep[$\sim$414\,pc,][]{Menten_2007} and nearly edge-on orientation, the Bar is a template to investigate the dynamics and chemistry in  strongly \mbox{FUV-irradiated} gas \citep[e.g.,][]{Pellegrini_2009,Cuadrado_2015a,Goicoechea_2016}. 

\section{Observations and data reduction}

The interferometric images were taken with the 7\,m antennas of the Atacama Compact Array (ACA), Chile. 
The observations consisted of a 10-pointing mosaic centered at \mbox{$\alpha$(2000) = 5$^h$35$^m$20.6$^s$};  \mbox{$\delta$(2000) = $-$05$^o$25$'$20$''$}. The total field-of-view (FoV) is $\sim$50$''$$\times$50$''$. Target line frequencies lie in the $\sim$345-358\,GHz range (Table~\ref{table_freqs}.1).
Lines were observed with correlators providing $\sim$500\,kHz resolution over a 937.5~MHz bandwidth. The ALMA-ACA observation time was $\sim$6\,h. 
In order to recover the  extended emission filtered out by the interferometer, we used fully sampled single-dish maps as zero- and short-spacings. The required high-sensitivity maps were obtained using the ALMA total-power 12\,m antennas ($\sim$19$''$ resolution). We used the GILDAS/MAPPING software  to create the short-spacing visibilities not sampled by ALMA-ACA. These visibilities were merged with the interferometric 
observations \citep[][]{Pety_2010}. 
The dirty image was deconvolved using the  H{\"o}gbom CLEAN algorithm. 
The resulting cubes were scaled from Jy/beam to brightness temperature scale using the synthesized beam size of 4.9$''$$\times$3.0$''$.
The achieved rms noise is $\sim$10-20~mK per 0.5~km~s$^{-1}$ smoothed channel, with an absolute flux accuracy of $\sim$10\%.
The resulting images are shown in Fig.~\ref{fig_figuron}.

In addition, we carried out pointed observations toward the DF with the  IRAM\,30\,m  telescope (Spain). The observed position lies roughly at the center of the ALMA-ACA field (see circles in Fig.~\ref{fig_figuron}). 
 This position is at $\Delta$RA$=$$+$3$''$ and $\Delta$Dec$=$$-$3$''$ from the  \mbox{``CO$^+$ peak''} of \citet{Stoerzer_1995}.  
 These multi-line observations are part of a complete
80-360\,GHz line survey at  resolutions 
between $\sim$30$''$ and $\sim$7$''$ \citep[][]{Cuadrado_2015a}.

\section{Results: morphology and emission properties}

In addition to the submillimeter (submm) emission images obtained with ALMA-ACA, panels b), c), and d) in Fig.~\ref{fig_figuron} show images of the DF traced by the vibrationally excited molecular hydrogen (H$_{2}^{*}$)  \mbox{v=1-0 $S$(1)} line \citep{Walmsley_2000}, of the atomic PDR (hydrogen is
predominantly in neutral atomic form) as traced by the \textit{Spitzer}/IRAC~8~$\mu$m emission \citep{Megeath_2012}, and of the ionization front (IF), the H/H$^+$ transition zone. The $\delta x$ axis shows the distance in arcsec to the IF.
Thus, in each panel, the FUV-photon flux decreases from right to left.

\begin{figure*}[t] 
\centering 
\includegraphics[scale=0.57,angle=0]{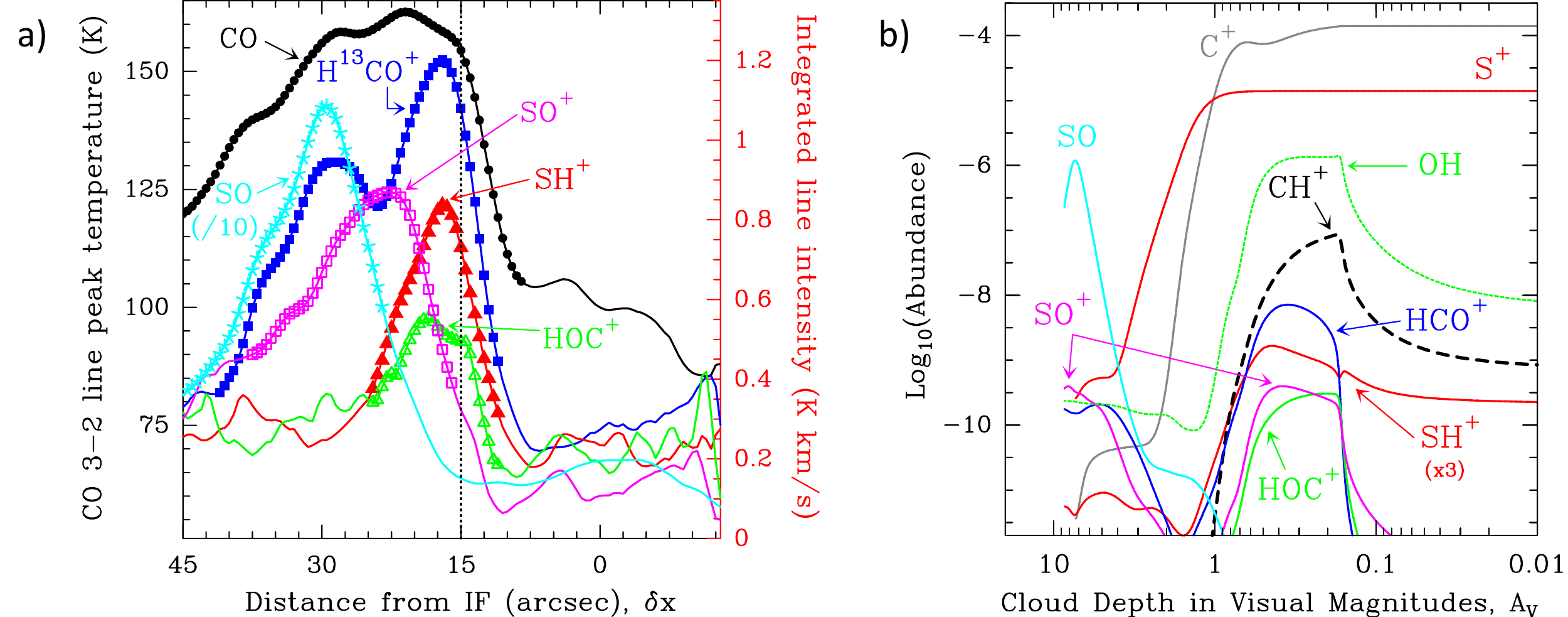}
\caption{a) ALMA-ACA averaged intensity cuts perpendicular to the Orion Bar. Except for \mbox{CO $J$=3-2} line (peak temperature in K), cuts show integrated line intensities (in K\,km\,s$^{-1}$). Symbols do not have any meaning, they are used to improve
Figure readability. 
b) Isobaric model of a PDR with 
$P_{\rm th}/k$=2$\times$10$^8$~K\,cm$^{-3}$ and $\chi$=10$^4$.  In both Figures, the FUV radiation from the Trapezium cluster comes from the right-hand side.} \label{fig_cuts}
\end{figure*}

The peak of the optically thick \mbox{CO~3-2} line provides a good lower limit to the gas temperature in the molecular PDR (\mbox{$\delta$$x$$>$15$''$}), with \mbox{$T_{\rm k} \gtrsim T_{\rm peak}^{\rm \,CO\,3-2}$$\simeq$150\,K} (Fig.~\ref{fig_figuron}a). 
The \mbox{ALMA-ACA} images show that, except for SO$^+$, the emission from reactive ions 
starts very close to the DF, and globally follows that of H$_{2}^{*}$. On small scales  ($\lesssim$3$''$$\approx$1000\,AU), several  emission peaks of these ions coincide with the brightest H$_{2}^{*}$ \mbox{v=1-0 $S$(1)}  peaks (e.g., at $\delta y \simeq -18''$). Although the SH$^+$ and HOC$^+$ peaks at $\delta y \simeq +27''$ do not exactly match a H$_{2}$ v=1-0 peak, observations do show the presence of extended H$_2$ \mbox{v=2-1} and 1-0 emission along the SH$^+$ and \mbox{HOC$^+$-emitting} zone \citep[][]{vanderWerf_1996}. In fact, H$_{2}^{*}$ emission from very high  vibrational levels (up to v=10 or $E$/$k$$\approx$50,000\,K) has recently been  reported \cite[][]{Kaplan_2017}.

To investigate the molecular emission stratification, in
\mbox{Fig.~\ref{fig_cuts}a} we show averaged emission cuts perpendicular to the Bar.
The  cuts demonstrate  that the \mbox{SH$^+$ 1$_0$-0$_1$} and \mbox{HOC$^+$~4-3} lines 
arise from a narrow emission layer (akin to a filament), with a half-power-width of 
$\Delta(\delta x)$$\simeq$10$''$ ($\simeq$0.02~pc), that delineates the DF.
The \mbox{H$^{13}$CO$^+$~4-3} line displays this emission peak close to the DF,
as well as another peak deeper inside the molecular cloud 
(at $\delta x$$\approx$30$''$) that is dominated by emission from 
the colder molecular cloud interior.
The  \mbox{SO~8$_9$-7$_8$} line arises from these more \mbox{FUV-shielded} zones. 
These  spatial emission trends  are supported by the different line-widths 
 (averaged over the ACA field of view, see Table~\ref{table_freqs}.1). In particular, the \mbox{SO~8$_9$-7$_8$} line
 is narrower ($\Delta$v=1.8$\pm$0.1~km\,s$^{-1}$) than the \mbox{SO$^+$ 15/2$^-$-13/2$^+$} ($\Delta$v=2.1$\pm$0.1~km\,s$^{-1}$),  
\mbox{HOC$^+$~4-3} ($\Delta$v=2.7$\pm$0.1~km\,s$^{-1}$), and
\mbox{SH$^+$ 1$_0$-0$_1$} ($\Delta$v=2.8$\pm$0.1~km\,s$^{-1}$) lines
that arise from the more FUV-irradiated gas near the molecular cloud edge.

We first derived H$^{13}$CO$^+$, HOC$^+$, SO$^+$, and SO 
rotational temperatures ($T_{\rm rot}$) and column densities ($N$) 
by building rotational population diagrams from our 
 \mbox{IRAM\,30\,m}  observations \citep[line survey position,][]{Cuadrado_2015a}. 
Results are shown in Tables \ref{Table_columns} and \ref{Table_results}.
H$^{13}$CO$^+$ and HOC$^+$ have high dipole moments 
(and SO$^+$ to a lesser extent, but see \mbox{Appendix~\ref{Appendix-montecarlo}}). 
Hence, the observed submm lines have moderate critical densities
(several 10$^6$~H$_2$\,cm$^{-3}$). The low-$J$ transitions toward the 
line survey position  are subthermal (\mbox{$T_{\rm rot}\approx15$\,K\,$\ll T_{\rm k}$}). Their column densities are
relatively small: from $\sim$10$^{11}$\,cm$^{-2}$ (assuming uniform beam filling), to $\sim$10$^{12}$\,cm$^{-2}$ (for a more realistic filamentary emission layer of
$\sim$10$''$ width  and correcting for beam dilution).

In addition, we estimated the average  physical conditions that lead to the H$^{13}$CO$^+$, HOC$^+$, SO$^+$ and SH$^+$ emission toward the line survey position. We used a  Monte Carlo model (Appendix~\ref{Appendix-montecarlo}) that includes inelastic collisions with H$_2$ and $e^-$, as well as radiative excitation by the far-IR
dust  radiation field in the region \citep[see][and Fig.~\ref{fig:continuum}]{Arab_2012}. This allowed us to refine the source-averaged column density estimation for a \mbox{10$''$--wide} emission layer  (Table~\ref{Table_columns}). 
The observed H$^{13}$CO$^+$, HOC$^+$, and SO$^+$ 
line intensities and $T_{\rm rot}$ are reproduced with \mbox{$n$(H$_2$)$\simeq$10$^5$\,cm$^{-3}$}, \mbox{$n(e^-)$=10\,cm$^{-3}$}, and 
\mbox{$T_{\rm k}$$\gtrsim$200\,K} (thus consistent with  
\mbox{$T_{\rm peak}^{\rm CO\,3-2}$)}. 
However, with the set of assumed SH$^+$ collisional rates, fitting the SH$^+$ lines requires denser gas, $\sim$10$^6$\,cm$^{-3}$ \cite[see also][]{Nagy_2013}. 
As $T_{\rm k}$ is expected to sharply vary along the DF \citep[e.g.,][]{Sternberg_1995}, the result of these single-component models should be taken as average conditions (over the ACA resolution). We note that at the distance to Orion, 4$''$ is equivalent to 1~mag of visual extinction for \mbox{$n_{\rm H}$=10$^5$\,cm$^{-3}$}.

Low [HCO$^+$]/[HOC$^+$] abundance ratios ($\sim$200-400) have previously  been inferred 
from lower angular resolution (10$''$ to 70$''$, depending on the line and telescope) pointed observations toward the Bar
\citep[][]{Apponi_1999,Fuente_2003}. 
Such ratios are much lower than those predicted  in FUV-shielded gas \citep[$>$1000, e.g.,][]{Goicoechea_2009}.
Given the similar $A_{\rm ul}$ coefficient and upper level energy ($E_{\rm u}$/$k$) of the \mbox{H$^{13}$CO$^+$} and \mbox{HOC$^+$ 4-3} transitions, their integrated intensity ratio is a good measure of the [HCO$^+$]/[HOC$^+$] isomeric ratio (lines are optically thin). The observed \mbox{H$^{13}$CO$^+$/HOC$^+$ 4-3} line ratio is equivalent to a roughly constant [HCO$^+$]/[HOC$^+$] ratio of 145$\pm$5 along the $\delta x$=12-22$''$ layer (assuming [HCO$^+$]/[H$^{13}$CO$^+$]=67). The  ratio then increases as the FUV-photon flux decreases, [HCO$^+$]/[HOC$^+$]=180$\pm$4 along the $\delta x$=22-32$''$ layer, until the \mbox{HOC$^+$ 4-3} signal vanishes deeper inside the molecular cloud.

\section{Discussion}

ALMA HCO$^+$ 4-3 images at $\sim$1$''$ resolution  suggest the presence of 
 high-pressure structures, 
\mbox{$P_{\rm th}$/$k$=$n_{\rm H}$\,$T_{\rm k}$$\approx$(1-2)$\times$10$^8$\,K\,cm$^{-3}$}, close to the
DF \citep{Goicoechea_2016}. 
Despite the different resolutions,  several of these HCO$^+$ peaks coincide with the brightest \mbox{SH$^+$ 1$_0$-0$_1$} and OH\,84\,$\mu$m peaks
(Figs.~\ref{fig_figuron}e and f). Thus, it is reasonable to assume
that SH$^+$ arises from these structures. 

We used version 1.5.2 of the Meudon PDR code \citep[][]{LePetit_2006} to model
the formation and destruction of reactive ions in a constant thermal-pressure slab of FUV-irradiated gas (\mbox{$P_{\rm th}$/$k$} from \mbox{2$\times$10$^7$} to \mbox{2$\times$10$^8$\,K\,cm$^{-3}$}).
Figure~\ref{fig_cuts}b shows abundance profiles (with respect to H nuclei)
predicted by the high-pressure model. The FUV-photon
flux  is $\chi$=10$^4$ (in Draine units), the expected radiation field close to the DF. Compared to \cite{Nagy_2013}, we have included  more recent rates for reactions of  S$^+$   with H$_2$(v) \citep{Zanchet_2013}. We adopt an undepleted sulfur abundance of [S]=1.4$\times$10$^{-5}$ \citep{Asplund_2005}.
 The predicted abundance stratification  in \mbox{Fig.~\ref{fig_cuts}b} qualitatively agrees with the observational intensity cut in  
 \mbox{Fig.~\ref{fig_cuts}a} (but recall that the spatial scales are not directly
 comparable, as for the studied range of \mbox{$P_{\rm th}$/$k$} in these 1D isobaric models, 10 mag of visual extinction corresponds to 1-10$''$). 
 To be more specific, we compared the inferred column densities with those predicted by the PDR model (\mbox{Table~\ref{Table_columns}}). 
Because the Orion Bar is not a perfect edge-on PDR, this comparison requires a knowledge of the tilt angle ($\alpha$) with respect to a pure edge-on geometry, and of the line-of-sight cloud-depth ($l_{\rm depth}$). 
Recent studies constrain $\alpha$ and $l_{\rm depth}$ to \mbox{$\lesssim$4$^o$} and $\simeq$0.28\,pc, respectively \citep{Salgado_2016,Andree_2017}. 
For this geometry, the intrinsic width of the SH$^+$- and \mbox{HOC$^+$-emitting} layer
would be narrower, from  $\simeq$10$''$ (the observed value) to $\simeq$2$''$ (if the Bar is actually tilted).
Given these  uncertainties, the agreement between the range of observed and predicted columns is satisfactory for HOC$^+$, H$^{13}$CO$^+$, and SO$^+$ (Table~\ref{Table_columns}).
Although PDR models with lower pressures predict qualitatively similar  stratification, and a reactive ion abundance peak  also at $A_{\rm V}$$\lesssim$1\,mag  (where  C$^+$, S$^+$, and H$_{2}^{*}$ coexist), a  model with ten times lower $P_{\rm th}$ underestimates
$N$(HOC$^+$) and $N$(SO$^+$)  by large factors ($\gtrsim$20).
This is related to the lower predicted OH column densities (by a factor of $\sim$25),  key in the formation of CO$^+$, HOC$^+$ and SO$^+$ \mbox{\citep[][]{Goicoechea_2011}}.
Interestingly, SO$^+$ peaks deeper inside the cloud, while in the PDR models the main SO$^+$ peak is close to that of SH$^+$.
Compared to the other ions, SO$^+$ destruction near the DF is dominated by  recombination and  photodissociation (not by reactions with H$_2$ or H). 
Hence, [SO$^+$] depends on the [OH]/$x_e$ ratio \citep[][]{Turner_1996} and on the photodissociation rate. The SO$^+$ line  peaks deeper inside the  cloud than the OH~$^2$$\Pi_{3/2}$ \mbox{$J$=7/2$^-$-5/2$^+$} line  (Fig.~\ref{fig_figuron}f) suggesting that near the DF either $n$($e^-$) is higher than in the model or, more likely, that the SO$^+$ photodissociation rate  is larger. Indeed, SO$^+$ can be dissociated by lower $E$ photons \citep[$<$5\,eV,][]{Bissantz_1992}. 
This process is not taken into account in the PDR model.

 \begin{table}[t]
\centering 
\caption{Inferred column densities and PDR model predictions.}     	
\begin{tabular}{c c c c @{\vrule height 10pt depth 2pt width 0pt}}   
\hline
\hline
    & log$_{10}\,N $ [cm$^{-2}$] & &  	 \\
	&	  Rotational Diagram & Non-LTE\tablefootmark{$\dagger$}  &	PDR 
	Model\tablefootmark{$\ddagger$}  \\    
\hline
H$^{13}$CO$^+$	&	  11.8\tablefootmark{a}-12.1\tablefootmark{b} & 12.3 &	11.0\tablefootmark{c}-12.2\tablefootmark{d}	\\
HOC$^+$			&	  11.6\tablefootmark{a}-11.9\tablefootmark{b} & 12.0 &	11.7\tablefootmark{c}-12.9\tablefootmark{d}	\\
SO$^+$	        &  	12.3\tablefootmark{a}-12.5\tablefootmark{b}   & 12.5 &	12.4\tablefootmark{c}-13.5\tablefootmark{d}	\\
SH$^+$	        &   	--    		                              & 13.2 &	11.7\tablefootmark{c}-12.8\tablefootmark{d}	\\
\hline
\end{tabular}
\tablefoot{
\tablefoottext{$\dagger$}{From a non-LTE  model.}
\tablefoottext{$\ddagger$}{Isobaric PDR model 
($P_{\rm th}/k$=2$\cdot$10$^8$\,K\,cm$^{-3}$ and $\chi$=10$^4$). 
Columns integrated up to A$_V$=10.}
\tablefoottext{a}{Assuming uniform beam filling.} 
\tablefoottext{b}{For a filament of 10$''$-width.} 
\tablefoottext{c}{Face-on geometry.}
\tablefoottext{d}{Edge-on geometry   with a tilt angle of \mbox{$\alpha\negmedspace\backsimeq$4$^{\circ}$.}}}
 \label{Table_columns} 
\end{table}

For SH$^+$, we infer a column density that is a factor \mbox{$\simeq$3-30}  above the PDR model prediction (depending on $\alpha$). 
Recall that reaction \mbox{S$^+$ + H$_2$(v) $\rightarrow$ SH$^+$ + H} (endothermic by \mbox{$\Delta E/k$$\simeq$9860\,K} when v=0) only becomes exothermic when v$\geq$2, but
remains slow even then \citep{Zanchet_2013}.
The mismatch between model and observation, if relevant, may suggest an additional source of SH$^+$  that is not well captured by the model: overabundant H$_2$(v$\geq$2) or temperature/pressure spikes due to the PDR dynamics \citep{Goicoechea_2016}.

Regardless of the excitation details, our observations show that detecting SH$^+$ and HOC$^+$ emission is an unambiguous indication of FUV-irradiated gas.
Intriguingly, \mbox{SH$^+$} and HOC$^+$  line-widths are narrower than those of CH$^+$  \citep[$\Delta$v$\simeq$4.5-5.5~km\,s$^{-1}$,][]{Nagy_2013,Parikka_2017}. The broader CH$^+$ lines  in the Bar have been interpreted as a signature of 
CH$^+$  high reactivity \citep{Nagy_2013}. In this view, the exothermicity  
(equivalent to an effective formation temperature of about 5360\,K)
of the dominant formation route, reaction \mbox{C$^+$ + H$_2$(v$\geq$1) $\rightarrow$ CH$^+$ + H}, goes into CH$^+$ excitation and motion. However, reactive collisions of CH$^+$ with H and H$_2$ are faster than elastic and  rotationally inelastic 
 collisions (see Appendices~\ref{Destruction-time-scales} and \ref{Appendix-montecarlo}). In particular, the CH$^+$ lifetime is so short 
 (a few hours)
that the molecule does not have time to thermalize, by elastic collisions, its translational motions to a velocity distribution at $T_{\rm k}$ \citep{Black_1998}.   Hence,  the broad  lines would be related to the energy excess upon
CH$^+$ formation (thousands of K) and not to the actual  $T_{\rm k}$ \citep[at 
\mbox{$T_{\rm k}$=1000\,K}, the CH$^+$ thermal line-width will only  be 1.8~km\,s$^{-1}$,][]{Nagy_2013}, 
nor to an enhanced gas turbulence.
Detailed models of the CH$^+$ excitation  show that inclusion of formation and destruction rates in the level population determination affects the high-$J$ levels \citep{Godard_2013}.
Indeed, using the CH$^+$ \mbox{$J$=3-2} to \mbox{6-5} intensities measured by 
\textit{Herschel}/PACS toward the line survey position \citep[][]{Nagy_2013}, we derive $T_{\rm rot}^{\rm PACS}$$\simeq$150\,K. This is significantly warmer than
 $T_{\rm rot}$ of  SH$^+$, HOC$^+$ and SO$^+$.

CH$^+$ is a light hydride \citep[see][for a review]{Gerin_2016} meaning that its lowest rotational transitions lie at far-IR  frequencies. Therefore, its critical densities are very high (\mbox{$n_{\rm cr}$(CH$^+$ 4-3)$\simeq$7$\times10^9$~H$_2$\,cm$^{-3}$}, cf. \mbox{$n_{\rm cr}$(HOC$^+$ 4-3)$\simeq$4$\times10^6$~H$_2$\,cm$^{-3}$} at $T_{\rm k}$=200\,K). Such values are much higher
than the  gas density in the Orion Bar. Thus, without an excitation mechanism 
other than collisions, one would expect $T_{\rm rot}$(CH$^+$)$\ll$$T_{\rm rot}$(HOC$^+$).
However, CH$^+$ has a shorter lifetime than the other ions, and shorter than the timescale for \mbox{inelastic} collisional excitations. Hence, the higher $T_{\rm rot}^{\rm PACS}$(CH$^+$)
compared to the heavier ions must be related to a formation, and perhaps radiative, pumping mechanism. In particular, the  dust continuum emission is much
stronger in the \mbox{far-IR}  than in the (sub)mm (Fig.~\ref{fig:continuum}). As a consequence, the \mbox{far-IR} CH$^+$ rotational transitions have larger radiative pumping rates 
than the (sub)mm transitions of the heavier ions. As an example, we derive \mbox{$B_{\rm 3\rightarrow4}\,U_{\rm 3\rightarrow4}^{\rm Dust+CMB}$\,(CH$^+$)/$B_{\rm 3\rightarrow4}\,U_{\rm 3\rightarrow4}^{\rm Dust+CMB}$\,(HOC$^+$)$\approx$100} 
(where $B_{\rm lu}$ is the stimulated absorption coefficient and $U$ is the energy density
produced by the dust emission and by the cosmic background; see Appendix~\ref{rates}).
Therefore, CH$^+$ can be excited by radiation many times during its short lifetime (and during its mean-free-time for elastic collisions) so that it can remain kinetically hot (large velocity dispersion) and rotationally warm while it emits. On the other hand, the heavier ions are driven relatively faster toward smaller velocity dispersion by elastic collisions (narrower lines) and toward lower $T_{\rm rot}$ by inelastic collisions.
For HOC$^+$, we estimate that several collisional excitations
take place during its lifetime (several hours).  
This subtle but important difference likely explains the  narrower lines and lower $T_{\rm rot}$ of the heavier reactive ions, as well as their slightly different spatial distribution compared to \mbox{CH$^+$}  (see Fig.~\ref{fig_figuron}e).

\begin{acknowledgements}

We thank  A. Parikka for sharing her \textit{Herschel} far-IR CH$^+$ and OH line maps.
We thank the ERC for funding support under grant \mbox{ERC-2013-Syg-610256-NANOCOSMOS},
and the Spanish MINECO under grants AYA2012-32032, AYA2016-75066-C2-(1/2)-P and CSD2009-00038.

\end{acknowledgements}


\bibliographystyle{aa}
\bibliography{references}

\begin{appendix}
  
\section{Spectroscopic parameters of the observed lines}

\begin{table*}[hb]
\label{table_freqs}
\caption{Frequencies, energy levels, Einstein coefficients and line parameters of the ALMA-ACA spectra averaged over the observed field.} 
\centering
\begin{tabular}{lcccccccc}
\hline\hline
               &                     &  Frequency  &  $E_{\rm u}/k$ & $A_{\rm ul}$ & $\int$$T$$d$v      &  v$_{\rm peak, LSR}$   &   $\Delta$v     & $T_{\rm peak}$ \\
Species        &  Transition         &    (MHz)    &       (K)      & (s$^{-1}$)   & (K\,km\,s$^{-1}$)  &    (km\,s$^{-1}$)      & (km\,s$^{-1}$)  &    (K) \\
\hline
CO             &    3-2              & 345795.9899 &  33.2  & 2.50$\cdot$10$^{-6}$ &  625.67 (0.72)      &   10.1 (0.1)        & 4.6 (0.1)     & 127.41 \\
HCO$^+$        &    4-3              & 356734.2246 &  42.8  & 3.55$\cdot$10$^{-3}$ &   33.93 (0.25)      &   10.5 (0.1)        & 3.1 (0.1)     &  10.41 \\
H$^{13}$CO$^+$ &    4-3              & 346998.3360 &  41.6  & 3.27$\cdot$10$^{-3}$ &    0.56 (0.01)      &   10.7 (0.1)        & 2.4 (0.1)     &   0.22 \\
HOC$^+$        &    4-3              & 357922.0140 &  42.9  & 1.82$\cdot$10$^{-3}$ &    0.21 (0.01)      &   10.6 (0.4)        & 2.8 (0.1)     &   0.07 \\
SH$^+$         & 1$_0$-0$_1$ 1/2-3/2 & 345944.4205 &  11.6  & 2.30$\cdot$10$^{-4}$ &    0.30 (0.01)      &   10.6 (0.2)        & 2.7 (0.1)     &   0.10 \\
SO$^+$         & 15/2$^-$-13/2$^+$           & 347740.0110 &  70.1  & 1.19$\cdot$10$^{-3}$ &    0.36 (0.01)      &   10.6 (0.1)        & 2.1 (0.1)     &   0.16 \\
SO             & 8$_9$-7$_8$         & 346528.5280 &  78.8  & 5.38$\cdot$10$^{-4}$ &    3.50 (0.08)      &   10.3 (0.2)        & 1.8 (0.1)     &   1.81 \\               
\hline
\end{tabular}
\tablefoot{Velocity-integrated line intensities, LSR velocity of the line emission peaks, line-widths and line temperature peaks obtained by a Gaussian fitting procedure.
 Parentheses indicate the fit uncertainty.} 
 \end{table*}


\section{Rotational population diagrams and column densities}

\begin{figure*}[hb]
\centering
\includegraphics[scale=0.5, angle=0]{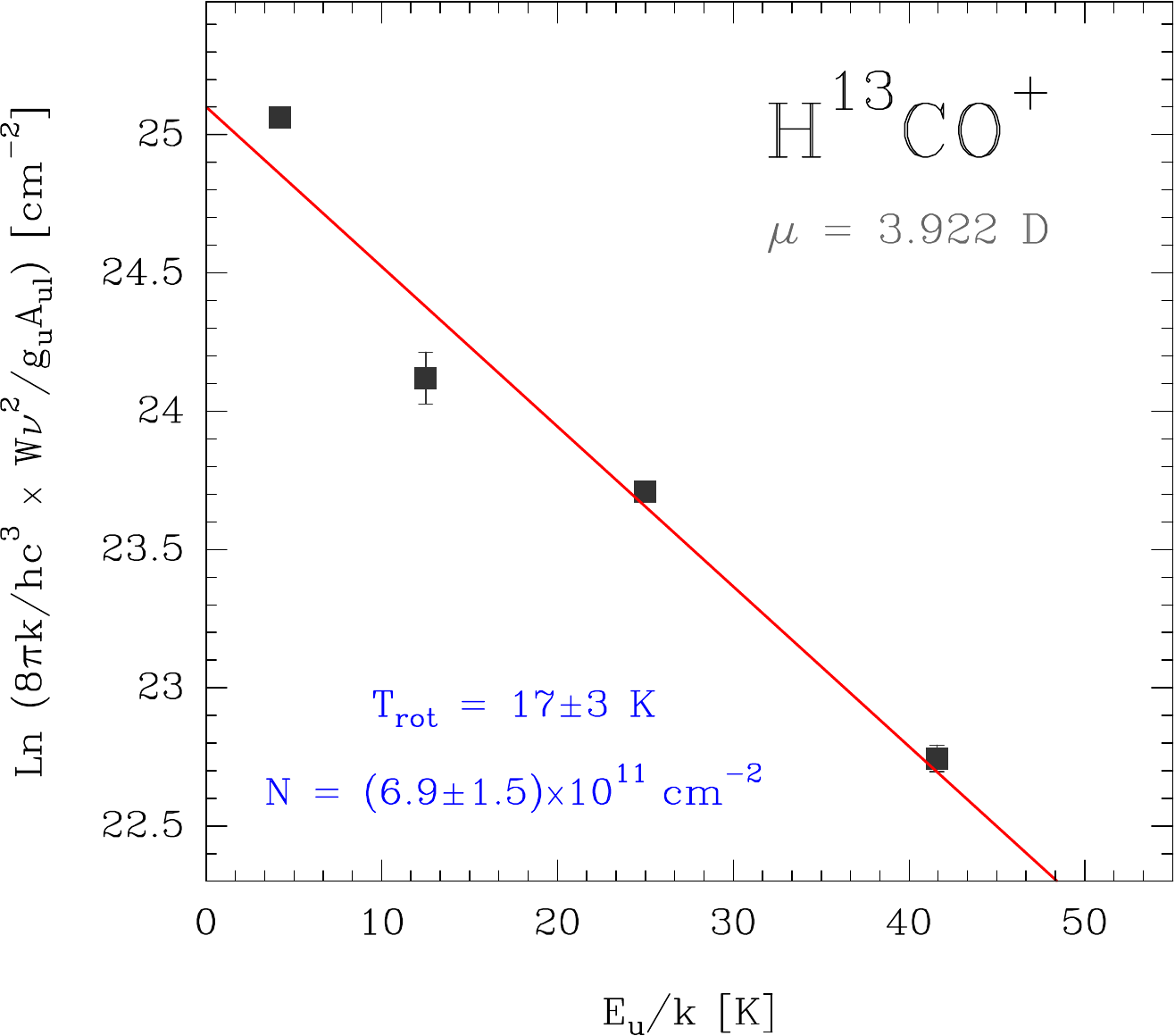} \hspace{0.5cm} 
\includegraphics[scale=0.5, angle=0]{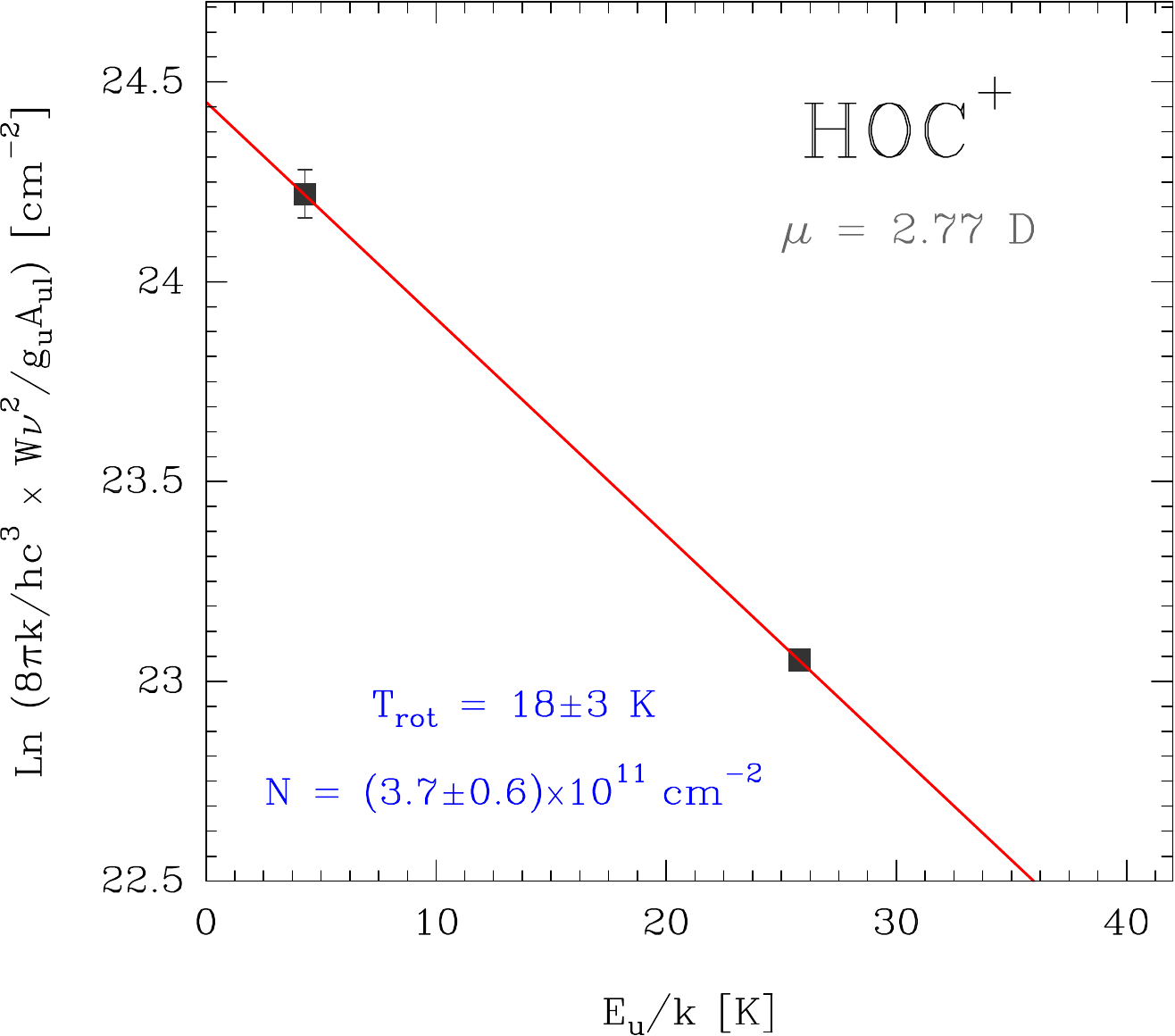} \\
\vspace{0.5cm}
\includegraphics[scale=0.5, angle=0]{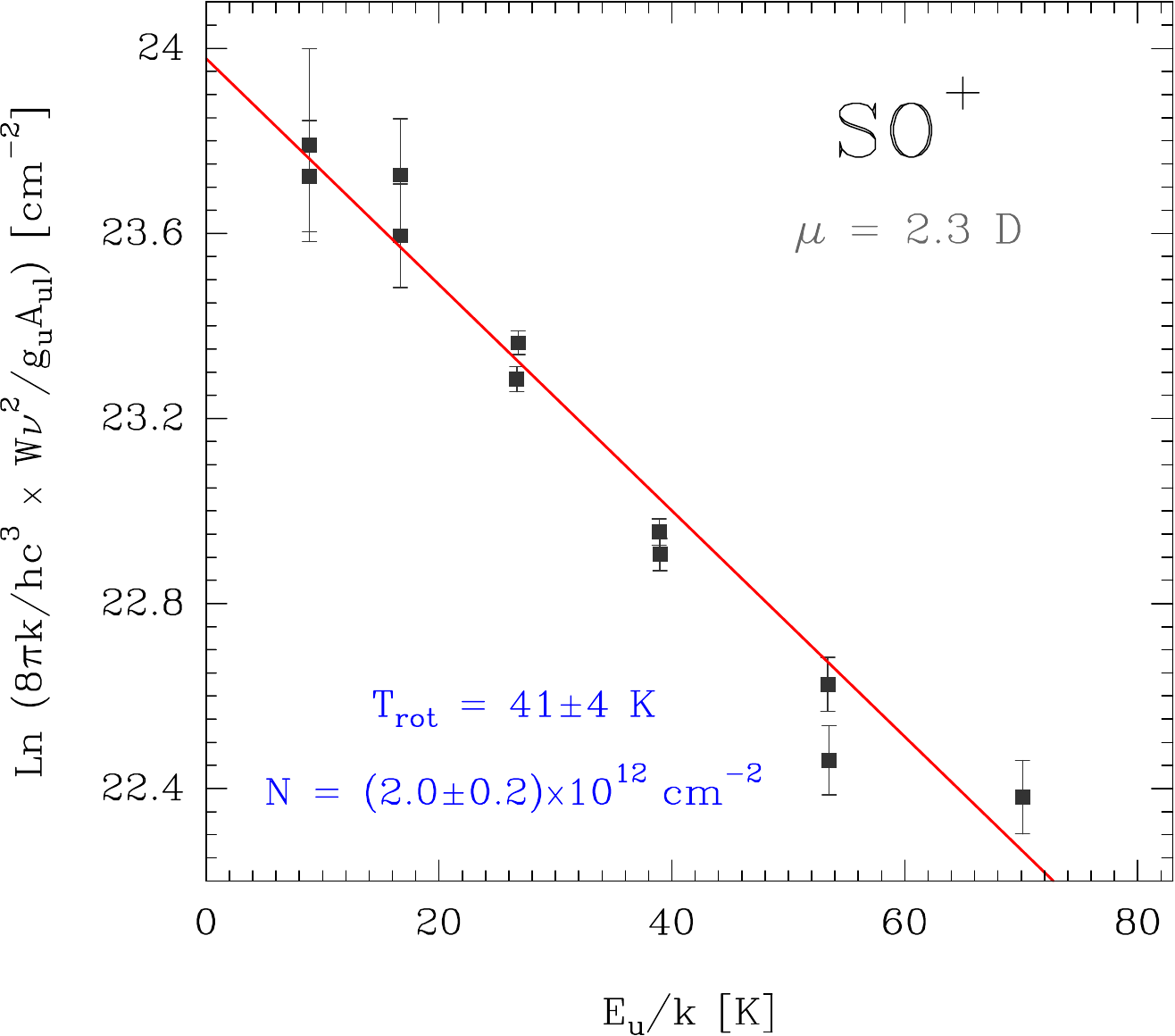} \hspace{0.5cm}
\includegraphics[scale=0.5, angle=0]{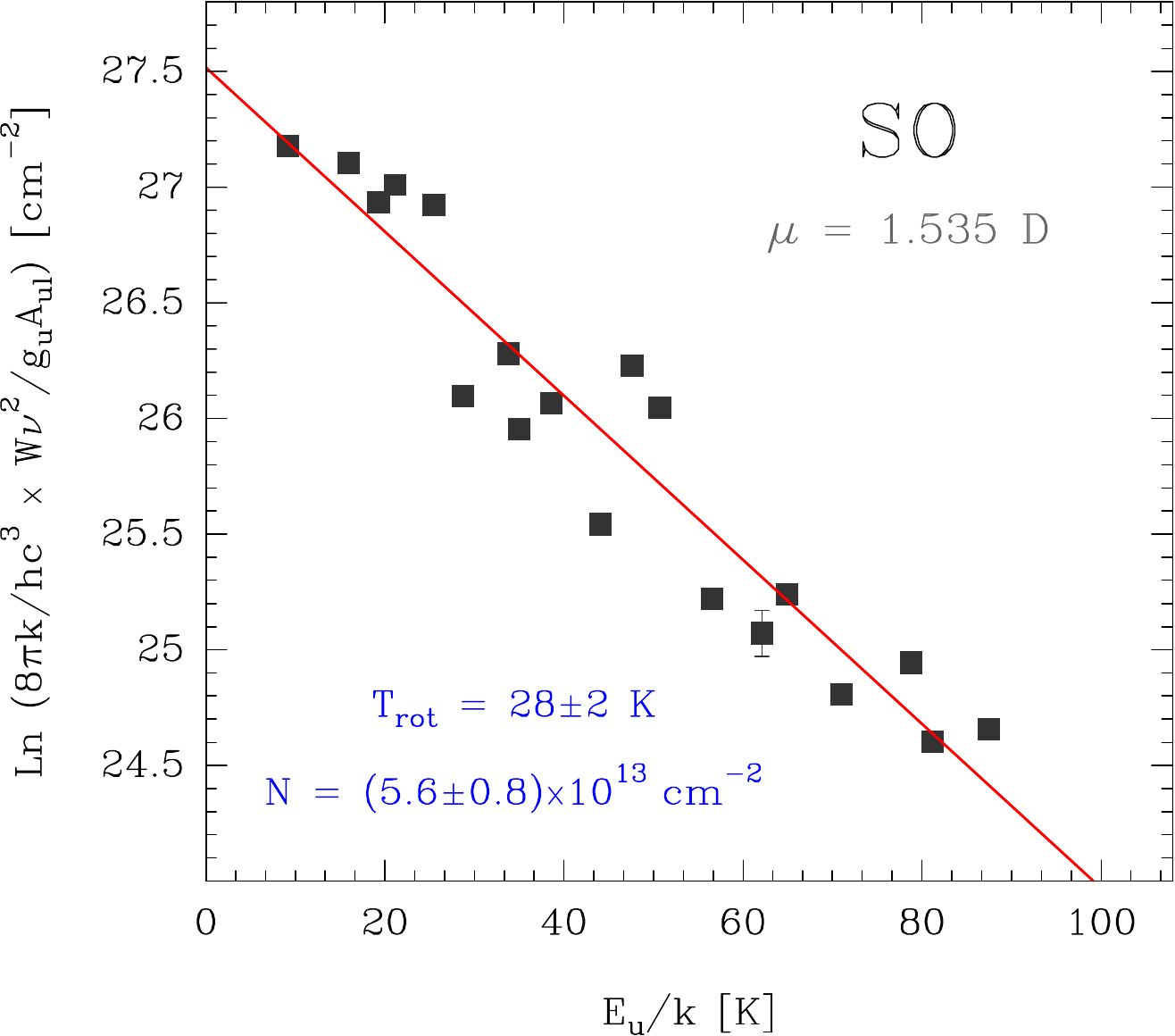} \\
\vspace{0.5cm}
\caption{Rotational diagrams obtained from IRAM\,30\,m observations toward the 
line survey position \citep{Cuadrado_2015a} assuming optically thin emission
and uniform beam filling. 
Fitted values of $T_{\rm rot}$, column density, and 
their respective uncertainties are indicated.}
\label{fig:DR}
\end{figure*}


\begin{table*}
 \centering 
 \caption{Rotational temperatures ($T_{\rm rot}$) and column densities ($N$) 
 inferred from IRAM\,30\,m observations toward the line survey position.}
 \label{Table_results}

 \begin{tabular}{l c c c c c @{\vrule height 10pt depth 5pt width 0pt}}     
 \hline\hline      
  & \multicolumn{2}{c}{uniform beam filling} \rule[0.15cm]{0cm}{0.2cm}\ &  &  \multicolumn{2}{c}{10$''$-width filament} \rule[0.2cm]{0cm}{0.2cm}\  \\ \cline{2-3} \cline{5-6}
  & $T_{\rm rot}$ & $N$(X) &  & $T_{\rm rot}$  & $N$(X) \ \\
  & $\mathrm{[K]}$ & $\mathrm{[cm^{-2}]}$ &   & $\mathrm{[K]}$ & $\mathrm{[cm^{-2}]}$  \\
\hline                                                                                                                                                                                                                                                          
H$^{13}$CO$^{+}$          &   17 $\pm$ 3     &   (6.9 $\pm$ 1.5) $\times$ 10$^{11}$  &  & 13 $\pm$ 2   &   (1.4 $\pm$ 0.5) $\times$ 10$^{12}$    \\              
HOC$^{+}$                 &   18 $\pm$ 3     &   (3.7 $\pm$ 0.6) $\times$ 10$^{11}$  &  & 11 $\pm$ 2   &   (7.8 $\pm$ 1.2) $\times$ 10$^{11}$    \\                                                                                                                                                                                 
SO$^+$                    &   41 $\pm$ 4     &   (2.0 $\pm$ 0.2) $\times$ 10$^{12}$   &   &    29 $\pm$ 3   &   (3.1 $\pm$ 0.5) $\times$ 10$^{12}$  \\    
SO                        &   28 $\pm$ 2     &   (5.6 $\pm$ 0.8) $\times$ 10$^{13}$  &  & 22 $\pm$ 2   &   (1.1 $\pm$ 0.2) $\times$ 10$^{14}$     \\    
SH$^{+}$                  &                  &                                       &  &  $\sim$10    &   (1.5 $\pm$ 0.5) $\times$ 10$^{13}$     \\   
\hline  
\end{tabular} 
 \end{table*}

\clearpage

\section{Chemical destruction timescales}
\label{Destruction-time-scales}

In this Appendix we use the rates of the chemical reactions  (included in our PDR model) that lead to the destruction of reactive ions, to compute their characteristic destruction timescales in the Orion Bar. For simplicity we adopt  $T_{\rm k}$=$T_{\rm e}$=300\,K.

The destruction timescales of
CH$^+$ by reactive collisions with H$_2$ and  H  are 
\mbox{$\tau_{\rm CH^+}(\rm H_2)$$\simeq$4.6~h (10$^5$\,cm$^{-3}$/$n_{\rm H}$)\,$f_{\rm H_2}^{-1}$} and 
  \mbox{$\tau_{\rm CH^+}(\rm H)$$\simeq$3.7~h (10$^5$\,cm$^{-3}$/$n_{\rm H}$)\,$(1-f_{\rm H_2})^{-1}$}, respectively. 
In these relations, $f_{\rm H_2}$$=$2$n$(H$_2$)/[$n$(H)+2$n$(H$_2$)] is the molecular gas fraction ($<$1 close to the DF of PDRs). 
The CH$^+$ destruction timescale by dissociative recombination
  is slower,   \mbox{$\tau_{\rm CH^+}(e^-)$$\simeq$185~h (10$^5$\,cm$^{-3}$/$n_{\rm H}$)\,(10$^{-4}$/$x_{e}$)}.
  
On the other hand, SH$^+$ slowly reacts with H$_2$ (the
  reaction is very endothermic). The relevant destruction timescales are  
 \mbox{$\tau_{\rm SH^+}(\rm H)$$\simeq$25~h (10$^5$\,cm$^{-3}$/$n_{\rm H}$)\,$(1-f_{\rm H_2})^{-1}$}  and
  \mbox{$\tau_{\rm SH^+}(e^-)$$\simeq$111~h (10$^5$\,cm$^{-3}$/$n_{\rm H}$)\,(10$^{-4}$/$x_{e}$)}. 
  
  For HOC$^+$ we derive
\mbox{$\tau_{\rm HOC^+}(\rm H_2)$$\simeq$14.6~h (10$^5$\,cm$^{-3}$/$n_{\rm H}$)\,$f_{\rm H_2}^{-1}$}  and   \mbox{$\tau_{\rm HOC^+}(e^-)$$\simeq$252~h (10$^5$\,cm$^{-3}$/$n_{\rm H}$)\,(10$^{-4}$/$x_{e}$)}.

SO$^+$ lifetime is longer. The reaction of SO$^+$ with H is very endothermic.
Close to the DF, SO$^+$ destruction is dominated  by dissociative recombination, with \mbox{$\tau_{\rm SO^+}(e^-)$$\simeq$139~h (10$^5$\,cm$^{-3}$/$n_{\rm H}$)\,(10$^{-4}$/$x_{e}$)}, and by photodissociation, with
\mbox{$\tau_{\rm SO^+}(\rm ph)$=27.8~h $e^{+1.7\,A_{\rm V}}$} for $\chi$=10$^4$.
In our PDR models we assume \mbox{$\kappa_{\rm SO^+}(\rm ph) = \chi \cdot 10^{-9}\,e^{-1.7\,A_{\rm V}}$~(s$^{-1}$)}, which is 
$\simeq$1000, $\simeq$10, $\simeq$5, and $\simeq$4 times faster than the adopted
$\kappa_{\rm HCO^+}$(ph), $\kappa_{\rm CH^+}$(ph), $\kappa_{\rm OH}$(ph), and $\kappa_{\rm SH^+}$(ph) photodissociation rates, respectively (at $A_{\rm V}$=1).
A summary of the relevant destruction timescales is presented in Table~\ref{table_Drates}.

\clearpage 
 
\section{Collisional and radiative excitation of H$^{13}$CO$^+$, HOC$^+$, SH$^+$, and SO$^+$}\label{Appendix-montecarlo}

To estimate the physical conditions of the reactive ions 
emitting gas, we solved the statistical equilibrium equations and radiative transfer using a non-local-thermodynamic equilibrium code 
\citep[][]{Goicoechea_2006}. In PDRs, the high electron density
(up to $n(e^-) \simeq x({\rm C^+})\,n_{\rm H} \simeq 10^{-4}\,n_{\rm H}$
for standard cosmic ray ionization rates)
 plays an important role in the collisional excitation of molecular ions 
and competes with collisions with H$_2$ and H.
This is because the associated collisional excitation rate coefficients
$\gamma_{\rm lu}(e)$ (cm$^3$\,s$^{-1}$) can be large, up to 10$^4$ times  $\gamma_{\rm lu}({\rm H_2})$,
and thus H$_2$ and $e$ collisional rates  (s$^{-1}$) become comparable if the ionization fraction $n(e^-)/n_{\rm H}$ is high, \mbox{$x_e$$\simeq$10$^{-5}$-10$^{-4}$}. 
Here we used published (or estimated by us)
 inelastic collisional rate coefficients\footnote{-- For H$^{13}$CO$^+$ and HOC$^+$ ($^1\Sigma^-$ ground electronic states) we used HCO$^+$-H$_2$ de-excitation rates of \cite{Flower_1999}, and specific \mbox{HCO$^+$-$e^-$} and \mbox{HOC$^+$-$e^-$} de-excitation rates \citep[presented in][and references therein]{Fuente_2008}. We computed the respective collisional excitation rates through detailed balance.\\ 
-- For SH$^+$ ($^3\Sigma^-$) there are no published collisional rate coefficients.
For SH$^+$-H$_2$ we simply scaled radiative rates (J.~Black, priv. comm.) for a \mbox{10-5000\,K} temperature range. For SH$^+$-$e^-$ collisions we used rate coefficients calculated in the
Coulomb-Born approximation for the 10-1000\,K temperature range (J.~Black, priv. comm.).\\
-- For SO$^+$ ($^2\Pi$) there are no published collisional rates.
Here we  modeled rotational levels in the $^2\Pi_{1/2}$
ladder only (the $^2\Pi_{3/2}$ ladder lies 525\,K above the ground, much higher than
the $T_{\rm rot}$(SO$^+$) inferred from our observations) and neglected
$\Lambda$-doubling transitions. For  SO$^+$-H$_2$, we used 
CS-He rates of \citet{Lique_2006} (CS has a similar molecular weight and dipole moment: 44 and 2.0\,D,
compared to SO$^+$: 48 and 2.3\,D. This is the adopted SO$^+$ dipole moment after Turner~1996). Rate coefficients were multiplied by 2 to account for the ionic character of SO$^+$.
For SO$^+$-$e^-$, we used rate coefficients calculated in the
Coulomb-Born approximation for the 10-1000\,K range (J.~Black, priv. comm.).\\},
and we adopted $n(e^-)$=10\,cm$^{-3}$ and \mbox{$n_{\rm H}$$\simeq$10$^4$-10$^6$\,cm$^{-3}$}
in the models,
the expected range of densities in the region. 

In addition to inelastic collisions, we included radiative excitation by absorption
of the 2.7\,K cosmic background and  by the (external) dust radiation field   in the region. The latter is modelled as a modified blackbody emission with an effective dust temperature of $T_{\rm d}$=50\,K, spectral emissivity index of $\beta$=1.6, and a dust opacity ($\tau_{\rm d}$)
of $\simeq$0.03 at a reference wavelength of  $\lambda$=160\,$\mu$m. The resulting 
continuum levels (Fig.~\ref{fig:continuum}) reproduce the far-IR photometric measurements toward the Orion Bar; we refer to 
\citet[][]{Arab_2012}. These authors estimated $T_{\rm d}$$\simeq$50\,K inside the Bar and $T_{\rm d}$$\simeq$70\,K immediately in front. 
Our calculations include thermal, turbulent, and line opacity broadening. 
The non-thermal velocity dispersion that reproduces the observed line-widths is 
$\sigma \simeq1$\,km\,s$^{-1}$ (with full width at half maximum of 2.355$\cdot$$\sigma$).

By varying $N$, $n_{\rm H}$ and $T_{\rm k}$, we tried to reproduce the H$^{13}$CO$^+$, HOC$^+$, SH$^+$, and  SO$^+$  line intensities observed by ALMA-ACA toward the line survey position as well as the intensities of the other rotational lines detected with the \mbox{IRAM-30m} telescope (correcting them by a dilution factor
 that takes into account the beam size at each frequency and an intrinsic \mbox{10$''$-wide} Gaussian filament shape of the emission).
Our best models fit  the intensities by less than a factor of 2. They also reproduce ($\pm$3\,K)
the $T_{\rm rot}$  inferred from the rotational population  diagrams 
for a  \mbox{10$''$-wide}  filament (see Table~\ref{Table_results}). 
For  H$^{13}$CO$^+$, HOC$^+$, and  SO$^+$, we obtain $n$(H$_2$)$\simeq$10$^5$\,cm$^{-3}$,
$n(e^-)$=10\,cm$^{-3}$, and $T$$_{\rm k}$$\gtrsim$200\,K. These conditions agree
with those inferred by \citet{Nagy_2017}  for the H$^{12}$CO$^+$ rotationally excited emission 
($J$=6-5 to 11-10) observed by \textit{Herschel}/HIFI.

For SH$^+$,  we 
reproduce the line intensities observed by ALMA-ACA at $\sim$345\,GHz, and by \textit{Herschel}/HIFI at $\sim$526\,GHz \citep[][]{Nagy_2013} (significantly diluted in the large HIFI beam according to our ALMA-ACA images), if the gas is an order of magnitude denser, $\sim$10$^6$\,cm$^{-3}$ (similar to \citet{Nagy_2013} as they
used the same estimated collisional rates).

\subsection{Radiative pumping rates (CH$^+$ vs. HOC$^+$)}
\label{rates}

\begin{figure}[hb]
\centering
\includegraphics[scale=0.5, angle=0]{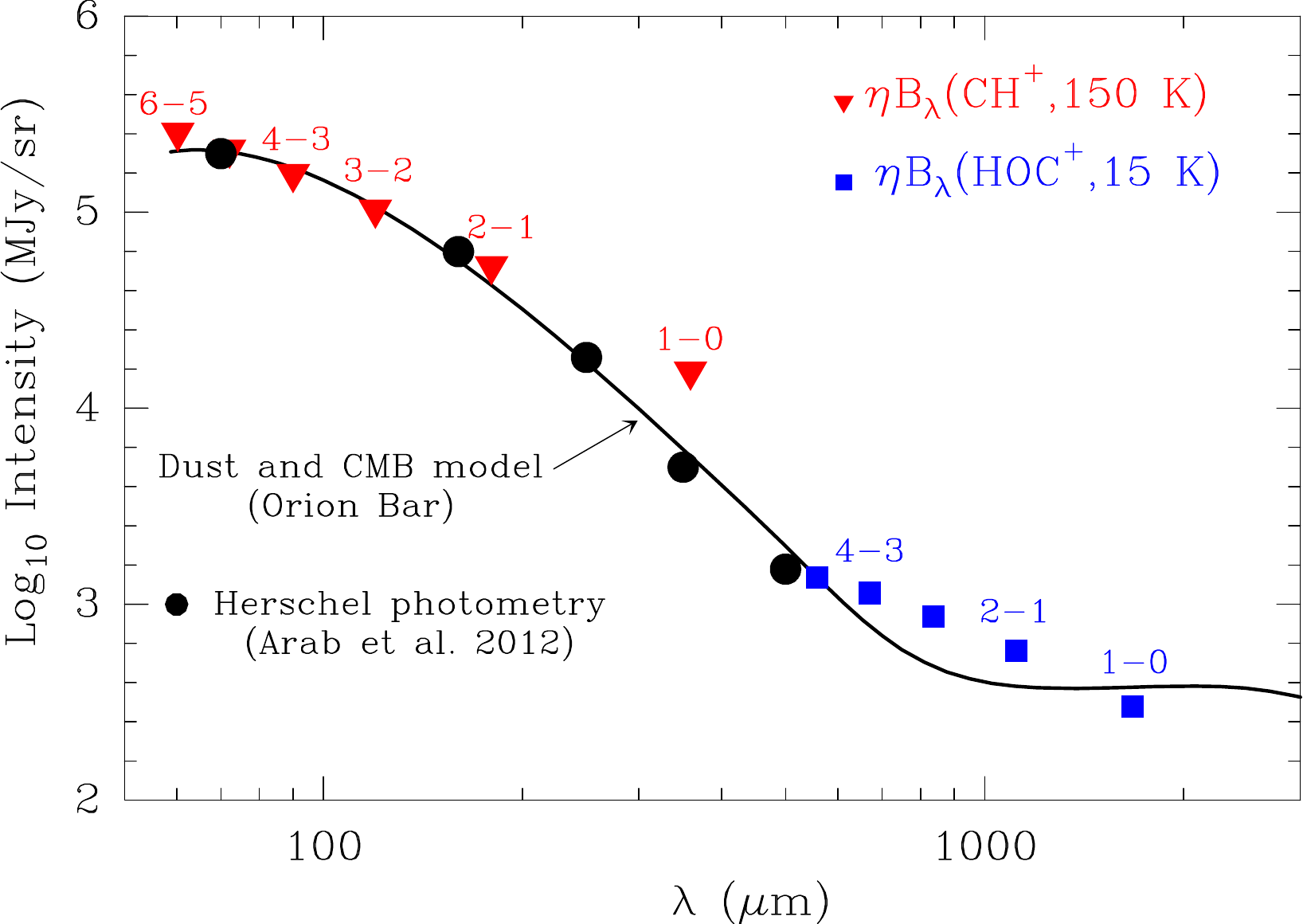} 
\caption{Dust continuum emission (with $T_{\rm d}$$=$50\,K) and cosmic millimeter background model used in the excitation calculation. Black circles show \textit{Herschel}'s photometric observations toward the Orion Bar. Red triangles show the wavelength position of
CH$^+$ rotational lines. Their intensity scale is that of a black body at 
150\,K (\mbox{$\approx$$T_{\rm rot}^{\rm PACS}$}) scaled by an arbitrary dilution factor $\eta$. 
Blue squares show the position of HOC$^+$ rotational lines.
Their intensity scale is that of a black body at 15\,K 
\mbox{($\approx$$T_{\rm rot}^{\rm IRAM\,30m}$)} scaled by a different arbitrary dilution factor.}
\label{fig:continuum}
\end{figure}

To support our interpretation, here we compare the collisional and radiative
pumping rates of  CH$^+$ and HOC$^+$ rotational lines with their chemical destruction
timescales. 
The inelastic collisional excitation rate ($C_{\rm lu}$) is given by
\begin{equation}
C_{\rm lu} = n({\rm H_2}) \, \gamma_{\rm lu}({\rm H_2}) +  n(e^-) \, \gamma_{\rm lu}({\rm {e^-}}) \,\,\,\,\,\,\, [{\rm s^{-1}}],
\end{equation}
where we assume that H$_2$ and $e^-$ are the only collisional partners. The upward excitation collisional coefficients $\gamma_{\rm lu}$ (cm$^{3}$\,s$^{-1}$) are computed from the
de-excitation coefficients by detailed balance
\begin{equation}
\gamma_{\rm lu} = \gamma_{\rm ul}\,\frac{g_u}{g_l}\,e^{-T^*/T_{\rm k}}
\,\,\,\,\,\,\, [{\rm cm^3\,s^{-1}}],
\end{equation}
where $T^* = h\nu/k$ is the equivalent temperature at the frequency $\nu$ of the transition. 
The continuum energy density in the cloud at a given frequency is 
\begin{equation}
U^{\rm Dust+CMB} = \beta \, [U(T_{\rm d}) + U(T_{\rm cmb})],  
\end{equation}
where $U(T_{\rm d})$ is the contribution from the external dust radiation field,
and $U(T_{\rm cmb}=2.73\,{\rm K})$ is the cosmic background. $\beta$ is the photon escape probability that tends to 1 if the line opacity tends to 0.
The radiative pumping rate can be written as
\begin{equation}
\label{rate-BU}
B_{\rm lu}\, U^{\rm Dust+CMB} = \beta \, A_{\rm ul} \,
\left[\frac{1-e^{-\tau_{\rm d}}}{e^{T^*/T_{\rm d}}-1} + \frac{1}{e^{T^*/T_{\rm cmb}}-1} \right]
\,\,\,\,\,\,\, [{\rm s^{-1}}],
\end{equation}
where $B_{\rm lu}$ and $A_{\rm ul}$ are the Einstein coefficients for stimulated absorption and for spontaneous emission, respectively. 
As an example of the excitation differences between CH$^+$ and the heavier reactive ions, we note that the CH$^+$ 4-3 line lies at a far-IR wavelength ($\lambda$$\simeq$90\,$\mu$m) where the 
intensity of the  continuum emission toward the Orion Bar is $\gtrsim$200 times stronger than that at the submm HOC$^+$ 4-3 line wavelength ($\lambda$$\simeq$837\,$\mu$m; see Fig.~\ref{fig:continuum}).

We adopt $n$(H$_2$)=10$^5$\,cm$^{-3}$, $n$($e^-$)=10\,cm$^{-3}$, 
\mbox{$T_{\rm k}$=$T_{\rm e}$=200\,K}, and our model of the
continuum emission (Fig.~\ref{fig:continuum}).
With these parameters, using the appropriate collisional rate coefficients \citep[see][for detailed CH$^+$ excitation models]{Nagy_2013, Godard_2013}, and 
in the $\beta$$=$1 limit, we compute the following collisional and radiative pumping rates 
\begin{equation}
\label{eq-col_chp01}
C_{\rm 01}\,({\rm CH^+}) = 3.7 \times 10^{-5} \,\, {\rm s^{-1}}
\,\,\,\,\, (\tau\simeq 7.6\,{\rm h}),
\end{equation}
\begin{equation}
\label{eq-rad_chp01}
B_{\rm 01}\, U^{\rm Dust+CMB} \,({\rm CH^+})= 1.3 \times 10^{-4}  \,\, {\rm s^{-1}}
\,\,\,\,\, (\tau\simeq 2.2\,{\rm h}),
\end{equation}
\begin{equation}
\label{eq-col_chp34}
C_{\rm 34}\,({\rm CH^+}) = 5.3 \times 10^{-6} \,\, {\rm s^{-1}}
\,\,\,\,\, (\tau\simeq 52.7\,{\rm h}),
\end{equation}
\begin{equation}
\label{eq-rad_chp34}
B_{\rm 34}\, U^{\rm Dust+CMB} \,({\rm CH^+})= 2.1 \times 10^{-3}  \,\, {\rm s^{-1}}
\,\,\,\,\, (\tau\simeq 0.1\,{\rm h}),
\end{equation}
for CH$^+$, and
\begin{equation}
\label{eq-col_hocp01}
C_{\rm 01}\,({\rm HOC^+}) = 1.2 \times 10^{-4}  \,\, {\rm s^{-1}} 
\,\,\,\,\, (\tau\simeq 2.3\,{\rm h}),
\end{equation}
\begin{equation}
\label{eq-rad_hocp01}
B_{\rm 01}\, U^{\rm Dust+CMB} \,({\rm HOC^+}) = 1.7 \times 10^{-5}  \,\, {\rm s^{-1}}
\,\,\,\,\, (\tau\simeq 16.4\,{\rm h}),
\end{equation}
\begin{equation}
\label{eq-col_hocp34}
C_{\rm 34}\,({\rm HOC^+}) = 7.9 \times 10^{-5}  \,\, {\rm s^{-1}} 
\,\,\,\,\, (\tau\simeq 3.5\,{\rm h}),
\end{equation}
\begin{equation}
\label{eq-rad_hocp34}
B_{\rm 34}\, U^{\rm Dust+CMB} \,({\rm HOC^+}) = 1.6 \times 10^{-5}  \,\, {\rm s^{-1}}
\,\,\,\,\, (\tau\simeq 16.9\,{\rm h}),
\end{equation}
for HOC$^+$. The quantities in parenthesis ($\tau$) represent the corresponding timescale for each excitation process \mbox{(see Table~D.2)}. In the case of reactive molecular ions, these rates compete with their chemical destruction timescales (derived in Appendix~\ref{Destruction-time-scales}).
Adopting a gas density of 10$^5$\,cm$^{-3}$, $f_{\rm H_2}$=0.5 and $x_e$=10$^{-4}$
for the Orion Bar,  the total CH$^+$ and HOC$^+$ chemical destruction timescales  are \mbox{$\tau_{\rm D}$(CH$^+$)$\simeq$4\,h} and \mbox{$\tau_{\rm D}$(HOC$^+$)$\simeq$26\,h}   (see \mbox{Table~\ref{table_Drates}}).
Comparing $\tau_{\rm D}$(CH$^+$) with the timescales for collisional and
radiative excitation shows 
that  CH$^+$ molecules are excited by radiation many times during its short lifetime,
but not by collisions. 
Hence, CH$^+$ can remain rotationally warm while it emits. Interestingly, the far-IR CH$^+$ lines observed by PACS 
 \citep[\mbox{$J$=3-2} to \mbox{6-5},][]{Nagy_2013} follow roughly  the same functional shape, \mbox{$\propto B_{\lambda}$\,($T_{\rm rot}^{\rm PACS}$$\simeq$150\,K}), 
 of the warm dust continuum emission (with $T_{\rm d}$$\simeq$50-70\,K; Fig.~\ref{fig:continuum}). 
Because for CH$^+$ radiative processes are faster than collisional processes
(Table~D.2), CH$^+$ molecules might have equilibrated with the dust radiation field they absorb \mbox{\citep{Black_1998}}. However,  \citet{Godard_2013} conclude that the
high-$J$ CH$^+$ lines in the Bar are mostly driven by formation pumping; thus,
naturally producing
warm rotational temperatures.
On the other hand, comparing \mbox{$\tau_{\rm D}$(HOC$^+$)} 
with the representative timescales for collisional and radiative excitation
shows that HOC$^+$ molecules (and likely the other heavier ions as well) are excited by collisions several times during their lifetime. Inelastic collisions thus can drive their rotational populations to lower $T_{\rm rot}$ relatively fast.


\begin{table*}[h]
\label{table_Drates}
\caption{Timescales, in hours, for chemical destruction by reactive collisions with H$_2$, H, $e^-$ and by FUV photodissociation.} 
\centering
\begin{tabular}{lcccc}
\hline\hline
Ion    &                $\tau$(H$_2$)$^{a}$                          &  $\tau$(H)$^{a}$                                                &  $\tau$($e^-$)$^{a}$                                           & $\tau$(photodiss.)$^{b}$ \\ \hline
CH$^+$ & 4.6\,h\,(10$^5$\,cm$^{-3}$/n$_{\rm H}$)\,$f_{\rm H_2}^{-1}$   & 3.7\,h\,(10$^5$\,cm$^{-3}$/n$_{\rm H}$)\,$(1-f_{\rm H_2})$$^{-1}$ & 185\,h\,(10$^5$\,cm$^{-3}$/n$_{\rm H}$)\,(10$^{-4}$/$x_e$) & 84.2\,h\,$e^{+2.94\,A_V}$\\ 
\textbf{$\tau_{\rm D}$$\approx$4\,h$^{c}$} & & & & \\ \hline
HOC$^+$ & 14.6\,h\,(10$^5$\,cm$^{-3}$/n$_{\rm H}$)\,$f_{\rm H_2}^{-1}$ &  ---                                                            & 252\,h\,(10$^5$\,cm$^{-3}$/n$_{\rm H}$)\,(10$^{-4}$/$x_e$) & 5144\,h\,$e^{+3.32\,A_V}$\\ 
\textbf{$\tau_{\rm D}$$\approx$26\,h$^{c}$} & & & & \\ \hline
SH$^+$  & ---                                                        & 25\,h\,(10$^5$\,cm$^{-3}$/n$_{\rm H}$)\,$(1-f_{\rm H_2})$$^{-1}$  & 111\,h\,(10$^5$\,cm$^{-3}$/n$_{\rm H}$)\,(10$^{-4}$/$x_e$) & 111\,h\,$e^{+1.66\,A_V}$\\ 
\textbf{$\tau_{\rm D}$$\approx$46\,h$^{c}$} & & & & \\ \hline
SO$^+$ & ---          &  ---                                                                                                           & 139\,h\,(10$^5$\,cm$^{-3}$/n$_{\rm H}$)\,(10$^{-4}$/$x_e$) & 27.8\,h\,$e^{+1.70\,A_V}$\\ 
\textbf{$\tau_{\rm D}$$\approx$73\,h$^{c}$} & & & & \\ \hline
\end{tabular}
\tablefoot{$^a$Assuming $T_{\rm k}$=$T_e$=300 K. $^b$For a FUV-radiation field of $\chi$=10$^4$. $^c$Total destruction timescale ($\tau_{\rm D}$) at $A_V$=1 assuming n$_{\rm H}$=10$^5$\,cm$^{-3}$ and $f_{\rm H_2}$=0.5.
For CH$^+$ (HOC$^+$), $\tau_{\rm D}$ is shorter (longer) than the timescale for non-reactive collisions in  their low-lying rotational levels (see Table D.2.).} 
\end{table*}

\begin{table*}[hb]
\vspace{2cm} 
\label{table_excitation_rates}
\caption{Representative timescales for non-reactive inelastic collisions and for radiative excitations of CH$^+$ and HOC$^+$ low-lying  levels.}
\centering
\begin{tabular}{ccccc}
\hline\hline
Ion    &  $\tau_{0\rightarrow1}$(coll.)$^{a}$ & $\tau_{3\rightarrow4}$(coll.)$^{a}$ & $\tau_{0\rightarrow1}$(rad.)$^{b}$  &  $\tau_{3\rightarrow4}$(rad.)$^{b}$ \\ \hline
CH$^+$ $J_{\rm l}\rightarrow J_{\rm u}$  &      7.6\,h                            &  52.7\,h  							  &   2.2\,h  						      &   0.1\,h \\
HOC$^+$ $J_{\rm l}\rightarrow J_{\rm u}$ &      2.3\,h                            &  3.5\,h 						      &   16.4\,h 							  &   16.9\,h\\
\hline
\end{tabular}
\tablefoot{$^a$Adopting $n$(H$_2$)=10$^5$\,cm$^{-3}$ and $n$($e^-$)=10\,cm$^{-3}$. $^b$Using our model of the
continuum emission in the Orion Bar (see Fig.~\ref{fig:continuum}).} 
\end{table*}

\end{appendix}

\end{document}